\documentstyle[12pt]{article}
\begin{document}
\def\thebibliography#1{\section*{REFERENCES\markboth
 {REFERENCES}{REFERENCES}}\list
 {[\arabic{enumi}]}{\settowidth\labelwidth{[#1]}\leftmargin\labelwidth
 \advance\leftmargin\labelsep
 \usecounter{enumi}}
 \def\newblock{\hskip .11em plus .33em minus -.07em}
 \sloppy
 \sfcode`\.=1000\relax}
\let\endthebibliography=\endlist

\hoffset = -1truecm
\voffset = -2truecm


\title{\large\bf
The Dimensions Of Field Theory : From Particles To Strings      
}
\author{
{\normalsize\bf
A.N.Mitra \thanks{e.mail: (1) ganmitra@nde.vsnl.net.in ;
(2) anmitra@csec.ernet.in}
}\\
\normalsize 244 Tagore Park, Delhi-110009, India.} 
\date{31 August 1999}
\maketitle

\begin{abstract}

This is an editorial summary of the contents of a Book comprising a set of 
Articles by acknowledged experts dealing with the impact of Field Theory on  
major areas of physics (from elementary particles through condensed matter 
to strings), arranged subjectwise under six broad heads. The Book which  
emphasizes the conceptual, logical and formal aspects of the state of the art 
in these respective fields, carries a Foreword by Freeman Dyson, and is to 
be published by the Indian National Science Academy on the occasion of the 
International Mathematical Year 2000. The authors and full titles of all 
the Articles (33) are listed sequentially (in the order of their first 
appearance in the narration) under the bibliography at the end of this 
Summary, while a few of the individual articles to appear in the Book 
are already available on the LANL internet.     

\end{abstract} 
  
\newpage



\section{Birth, Decline And Rebirth Of Field Theory}

If one must choose one single item of Twentieth Century Physics which 
stands out by the yardstick of most pervasive and decisive influence on  
its total development, Quantum Field Theory (QFT) certainly wins hands down.  
Historically, QFT  was born out of the marriage of Relativity 
and Quantum Theory, at a hefty price of mathematical self-consistency 
underlying the celebrated Dirac Theory, whose full significance took 
several stages to unfold through the vissicitudes of logical deduction, 
Šgoing well beyond the immediate discovery of the positron. Indeed of far 
greater significance from the conceptual point of view, was the 
realization that the "sea of negative energy states" was already a tacit
admission of the failure of relativistic quantum mechanics of a single
particle, in favour of a collective many-particle, or $field$ 
description, a fact which was to be driven home by Dyson in his 
Cornell lectures of 1952. And once this realization dawned on the pioneers,
the Klein-Gordon theory of scalar particles found a natural place in the 
new scenario, at the hands of Pauli-Weisskopf(1934) who now found little
difficulty in quantizing these bosonic particles just as easily as the 
Dirac theory had done to fermions. Thus was born "Quantum Field Theory" 
(QFT) in its full glory, with  Anti-matter playing a symmetrical role 
to Matter, irrespective of its fermionic or bosonic nature. [ Feynman's 
brilliant positron theory was a bold attempt to resurrect the single 
particle quantum mechanics description via "zigzag" diagrams (negative 
time propagation of negative energy electrons), but the more universal 
language of Field Theory eventually carried the day].       
\par
	QFT registered its first major success in the Covariant formulation 
of QED at the hands of Tomonaga and Schwinger on the one hand, and Feynman
on the other, with Dyson playing the catalyst-role in synthesizing the two.
This theory, in the course of circumventing unphysical infinities in the
measurable quantities, gave rise to a new dogma of  $Renormalizability$
which was to act as the yardstick of acceptability of theories to come. 
This dogma, together with the independent principle  of "Gauge Invariance" 
(already in-built in QED), were to be two pillars of QFT in its march 
towards greater victories to come, especially in the formulation of
strong interaction theories on analogous lines to QED. This led to the 
Yang-Mills theory (1954) of $SU(2)$ gauge bosons as the non-abelian 
counterpart of QED. Yet the path of QFT for strong interactions was strewn 
with so many thorns that for a long time it simply refused to move. Indeed
at the meson-baryon level there was a (temporary) disenchantment with 
QFT, in favour of a new paradigm of Hadron Democracy (somewhat akin to the 
Mach Principle) based on a selection of items from the full QFT package viz., 
analyticity of S-matrix elements subject only to unitarity. This was the 
"Bootstrap" Philosophy of Chew which held its ground for a brief period, 
until the discovery of Quark substructures (1963-64) brought the Hadrons 
down from the pedestal of elementary particles to a (more modest) composite 
status, which told (by hindsight) the reason why the Yang-Mills gauge theory 
of strong interactions had not worked at the composite hadron level of 
gauge fields. But now the quarks and gluons offered a fresh basis for the 
application of the Gauge Principle at a deeper level of elementarity; thus 
was born the non-abelian QCD for strong interactions (1973). 

\subsection{QFT in Action}

	With the rebirth of QFT, three principal weapons in its arsenal 
that have stood the test of time may be listed as {\bf Renormalizability},
{\bf Gauge Principle}, and {\bf Spontaneous Symmetry Breaking}. Indeed  
these are the 3 pillars on which rests the grand edifice of Field Theory 
encompassing the diverse phenomena of Nature within an integrated structure.
In this entire development, {\bf Symmetry} has all along played the role of 
the "guiding light", with a conservation law (Noether) associated with a
specific symmetry (invariance) type (Lorentz, Gauge, Chiral), making use
of the powerful language of {\it Group Theory} and {\it Topology} at its 
command. And in reverse, Symmetry breaking (spontaneous or dynamical), with 
its universal appeal, has been a key element in the understanding of a whole 
range of phenomena from condensed matter to particle physics and early 
Universe cosmology, with definite experimental and observational support. 
\par
	With the evolution of further unifying principles like $duality$ 
and $Supersymmetry$, QFT has greatly consolidated its  grip on Particle 
Physics, and extended its frontiers beyond its traditional domains, to 
the new heights of String Theory to "rein in"  the formidable force of 
Gravity. These powerful weapons at the command of QFT have led to new 
insights which have helped reveal its hold on widely different branches of 
Physics (which had hitherto evolved on their own steam), a saga of victories
in which the techniques of Path Integrals, Renormalization Group (RG) 
theory and Dirac/Bargmann constrained dynamics have played key roles. 
Especially noteworthy is the unification of the QFT language  with those of 
Quantum Statistical Mechanics (QSM) and Condensed Matter Physics (CMP), 
leading all the way to Cosmology and Black Hole Physics (thanks to the 
unifying power of String Theory).  
\par
	Progress has been far from uniform in the different sectors of QFT. 
In the Particle Physics domain which represents the Ultimate Laboratory for
testing the most profound concepts emanating from QFT, the current state
of the art is symbolized by the Standard Model (SM) designed to unify 
the three gauge sectors of strong, electromagnetic, and weak interactions.
Of these, the last two sectors are unified by the Glashow-Salam-Weinberg
(GSW) model of $SU(2) \otimes U(1)$ comprising photons and weak bosons, 
while a proper understanding of the QCD sector for $SU(3)$ strong 
interaction, still remains a distant goal. In this regard, a partial success
has been achieved in the perturbative domain of asymptotic freedom, while
the non-perturbative domain of $Confinement$  still remains elusive. This 
is reminiscent of the Churchillian phrase of "so much effort being spent 
towards so little effect" that was often used for the two-nucleon problem 
in the fifties, leading Hans Bethe to invoke his famous "Second Principle 
Theory" for effective nuclear interaction, which now seems to have shifted 
to the quark level. A part of the Book is devoted to the Strong Interaction 
problem of Quark Confinement from several different angles. 
\par
	And yet the methodology and techniques QFT have shown a striking
capacity to handle the problems of widely different sectors of physics 
well beyond particle physics, without changing the thematic framework, often 
with great success. Examples are QSM and Toda FT for non-linear systems. 
\par
	Low dimensional Field Theories (2D, 3D) are often useful not only
as soluble models designed to throw light on interesting features of QFT
which often remain obscured in 4D form,  but also as actual prototypes for 
systems moving in lower dimensions; (e.g., 2D conformally invariant field 
theory adequately describes the long range behaviour of systems undergoing 
second order phase transitions).  A striking example of 2D QFT is the 
Schwinger Model (both chiral and non-chiral) which has been subjected to
deep scrutiny from several angles, each with rich dividends. Similarly
$(2+1)D$ QFT , the so-called Chern-Simons (CS) Theory, has found rich 
applications in Quantum Hall Effect in Condensed Matter Physics. And
most significantly from the point of view of formal QFT, Witten's 
demonstration of an exact connection of $(2+1)D$ QFT with Jones Polynomials
did fulfil the long-cherished goal of an exact (non-perturbative) solution 
of a gauge field theory, at least in 3 dimensions for the first time.
\subsection{Scope of The Book}

	This Book is an attempt to capture a cross section of the
multifaceted flavour of QFT that has evolved over this Century, by putting
together a collection (albeit subjective) of Articles by acknowledged experts 
in their respective fields. Admittedly, the selection is constrained by the 
accessibility of experts to $this$ Editor  within a relatively short span 
of time, and of necessity leaves out many important areas of QFT studies. 
The style of each Article varies from author to author, but the  emphasis 
by and large is on conceptual and logical aspects of QFT formalism to the 
topic under study, designed to be instructive for a fairly wide class of
readership, (with enough access to references for those who wish to pursue
a particular line further), while the actual details of QFT methodology, 
or applications to phenomenology are outside the scope of the Book. Its 
theme is defined through a subjectwise classification of its contents under 
the following heads, in accordance with the specific sector/sectors of 
physics intended for exposition:   
\par \noindent
A): Basic Srtucture of QFT ($RG$ Theory; $SM$; $SSB$; Confinement) \\
B): Topological Aspects of QFT  \\
C): Formal Methods in QFT (QSM; Toda FT; LF and Constrained Dynamics)  \\ 
D): Extension of QFT Frontiers (SUSY; CFT; String Theory)     \\      
E): QFT In $(2+1)D$: CS Theory and Applications      \\     	        
F): Methods of Strong Interaction in QFT  \\
G): Conclusion (Concerning Foundations of Quantum Theory)      
\par
In this Introduction to the Book, an attempt has been made to organize the
contents under these categories, with a section devoted to each. The 
topics are arranged in descending order from `classical' to `evolving',
with the former playing the background to the latter. The narrative draws
freely from the perspective and language provided by the Authors concerned,
often without quotes. The referencing, (except for a few special papers
which are cited in the running text), is left to the Articles concerned, 
whose authorships are listed in the order of their appearance in this
narrative, in the bibliography at the end.    

\section{Some Core Aspects Of QFT}

In this Part we collect the Articles relating successively to RG Theory;
Electroweak coupling in the Standard Model; the dynamics of Symmetry
Breaking in different sectors of physics; and $two$ novel mechanisms of
Confinement in QCD, one proposed by {\bf Gribov} and one by {\bf Nishijima}.
    
\subsection{RG Theory in QFT}
	
	Although the $Renormalization$ strategy had originated in QED in the
(limited) context of absorbing divergences in physical entities like mass
and coupling constant (charge), it turned out that the concept itself has a
deeper meaning with much wider ramifications, which was later to get 
formalized as Renormalization Group $(RG)$ Theory.  The perturbative $RG$ 
formulated by "Stuekelberg-Petermann in 1952-53 as a group of infinitesimal
transformations, related to finite arbitrariness arising in the S-matrix
elements upon elimination of the UV divergences" - (D.V.Shirkov [1]).
In a parallel development, Gell-Mann-Low (1954) derived functional equations
for the QED propagators in the UV limit, on the basis of Dyson's (1949)
renormalization transformations, but missed the `group character' implied
in these equations. Finally, Bogoliubov-Shirkov (1955-56) put both aspects
together and derived the "$RG$-equations" in a form which brings out the 
`scaling' properties of the electron and photon propagators. Thus $RG$
invariance boils down to the invariance of a solution w.r.t. the manner
of its parametrization. These equations were further developed and made
more rigorous with mathematicians and physicists working in tandem, so
that renormalization became a well-developed method at the computational
level. But the underlying physical concepts behind these equations took 
some more time to unfold until after Kadanov's, and especially Wilson's 
pioneering work on the understanding of the "critical indices" in phase
transitions brought out the real physics behind the $RG$ equations. 
\par
	Wilson's work revealed the rich applicational potential of the 
$RG$ ideas in various fields of physics, from `critical phenomena' 
(spin lattices, polymer theory, turbulence) in condensed matter physics, 
to QCD parameters like the strong coupling constant $\alpha_s$ and the 
`running mass' $m(p^2)$. In particular, the discovery of Asymptotic freedom
in QCD allowed physicists to produce a logically consistent picture of
renormalization, one in which the perturbative expansions at any high 
energy scale can be matched with one another, without any need to deal
with intermediate expansions in powers of a large coupling constant.
Another important aspect of these $RG$ equations which has been emphasized
by the Dubna School, is the concept of {\it functional self-similarity} 
in mathematical physics, which has led to applications like the study of 
strong non-linear regimes: asymptotic behaviour of systems described by 
non-linear partial differential equations; problem of generating higher                   
harmonics in plasmas, and so on. The Book begins with a perspective Article 
by {\bf Dmitri Shirkov}[1] on all aspects of the subject, from an introduction to 
$RG$ in QFT to an overview of its methodology, together with applications 
of $RG$ ideas in some important arenas of physics.     
\par
	A relatively new approach to  $RG$ theory, termed "Similarity 
Renormalization Group" (SRG) was launched in this decade by Wilson and 
Glazek, as well as Wegner, and is based on the perception that divergences 
are in the first place due to the{\it locality} of the primary interactions.
For a proper understanding of the features of the SRG theory, it is enough 
to consider only the non-relativistic quantum mechanics (the usual UV
divergences of relativistic QFT are not relevant here!), where the locality
condition on the potentials at all scales corresponds to taking only delta 
functions and their derivatives. The associated divergences can be regulated
by introducing cut-offs whose effects may be removed by renormalization. 
\par
	In the SRG, the transformations that explicitly "run" the cut-off
parameter are developed. These similarity transformations are of course 
unitary, and constitute the group elements of SRG. They are characterized 
by a "running" cut-off on energy differences (not states). If the 
Hamiltonian is viewed as a large matrix, these cut-offs limit the 
off-diagonal matrix elements, and as they are gradually reduced, the 
Hamiltonian is forced towards the diagonal form. The perturbation 
expansion of the transformed Hamiltonians contains no small energy 
denominators, so that the expansion does not break down unless the
strengths of the interactions themselves are large. With the help of an 
associated concept of {\it coupling coherence}, SRG acquires respectability
as a proper theory with the {\it same} number of parameters as the original
(fundamental) theory. A review of the formalism and working of SRG  is 
given by {\bf R J Perry}[2], using as an example the exactly soluble case 
of a simple 2D delta function to act as a laboratory for testing the 
convergence of the SRG method in some detail.  
     
\subsection{Standard Model And Electroweak Coupling} 
	      
	The Gauge Principle, as a central ingredient of QFT, needed to be
supplemented with fresh ideas and paradigms, within its broad framework,  
to extend its tentacles further. One such idea was based on the degenerate 
structure of the vacuum, dominated by vales and hills, which crystallized 
eventually as a new theme termed "Spontaneous Symmetry Breaking" $(SSB)$, 
together with its  companion  "Dynamical Breaking of Chiral Symmetry" 
$(DB\chi S)$, which would now enable gauge fields to acquire mass in a 
subtle but self-consistent manner. Armed with this paradigm, the Gauge 
Theory registered a signal success in the Weak interaction sector, 
culminating in the Glashow-Salam-Weinberg $(GSW)$ Model of Electro-weak
Interactions, which offered a unified view of weak and electromagnetic 
interactions in the form of an $SU(2) \otimes U(1)$ gauge theory. A more 
ambitious form of unification of the three principal gauge fields as a 
straightforward extension of the $GSW$, so as also to include the strong 
Šinteraction (QCD) sector, under the umbrella of "Grand Unification Theory" 
$(GUT)$, did not unfortunately bear fruit, so that, for the time being,
the "Standard Model" $(SM)$ has had to rest content with only a partial
unification $SU(3) \otimes SU(2) \otimes U(1)$ of these gauge fields. 
Nevertheless this episode brings out a truism about the unpredictability
of Nature, viz., its refusal to yield to a particular strategy for a second
time, merely on the strength of its success on a previous occasion.  
\par
	In a highly instructive and self-contained Article, {\bf V Novikov}
[3] gives a panoramic view of the conceptual and methodological framework of 
QFT (with the ingredients of gauge principle, renormalization group, and 
spontaneous symmetry breaking) that have been employed in the formulation 
of $SM$ for elementary particle physics. He dwells in particular on the
Higgs mechanism for the generation of the fermion masses for several
generations, and brings out the powers of "loop corrections" in $SM$
to predict accurate bounds on the masses of as yet undiscovered particles.   
This is vividly illustrated by the "correct" mass of the $t(op)$-quark
$ahead$ of its experimental discovery, stringent limits on the Higgs
mass from the "Landau pole" structure of the running coupling constant,
and the windows to the "physics beyond $SM$" that such analyses provide.

\subsubsection{Discrete Symmetries in SM}
   
An essential aspect of the Standard Model concerns the role of discrete
symmetries $P,C,T$ in determining the structure of the electroweak coupling.
This subject has had a long history since the original Lee-Yang discovery of
$P$-violation, going through successive phases of chiral symmetry
(Landau-Salam), $CP$ invariance (Lee-Oehme-Yang), its subsequent violation
(Cronin-Fitch), and {\it ipso facto (?)} $T$-violation, a topic of intense 
experimental activity today. [This last is of course an immediate
consequence of $TCP$-invariance (Pauli-Lueders Theorem), which puts the
existence of antiparticles exactly on par with particles]. A brief state-
of-the-art review of the subject by {\bf P K Kabir} [4] follows.    

\subsection{Dynamics Of Symmetry Breaking}

Just as "Symmetry dictates interactions"- (C.N.Yang at the First Asia 
Pacific Conf, Singapore, 1983), the dynamical effects of its $breaking$ 
(whether spontaneously or dynamically) during out-of-equilibrium phase 
transitions is equally at the root of a whole range of phenomena from 
condensed matter to particle physics, and so on, all the way to early 
universe cosmology. Indeed the dynamics of non-equilibrium phase 
transitions and the $ordering process$ that occurs until the system
reaches a broken symmetry equilibrium stage, have developed in tandem 
with controlled experimental techniques in many areas of condensed matter
physics (binary fluids, ferromagnets, superfluids, liquid crystals), so
as to provide a solid basis for describing the dynamics of phase ordering.
In cosmology, measurements of Cosmic Microwave Background anisotropies, 
and the formation of large scale structures in the Universe, provide
signatures for phase transitions during and after $inflation$. And at
the accelerator energies (Brookhaven-RHIC or CERN-LHC), phase transitions
predicted by QCD could occur out of equilibrium via pion condensates. 
\par
	In an instructive review on this subject, {\bf Boyanovsky and
de Vega} [5] describe the relevant aspects of the dynamics of symmetry
breaking in many areas of physics (from condensed matter to cosmology)
vis-a-vis possible experimental signatures. In condensed matter, they 
address the dynamics of phase ordering, emergence of condensates, and
dynamical scaling. In QCD, the possibility of disoriented chiral pion 
condensates arising from out-of-equilibrium phase transitions is considered.
And in the early Universe, the dynamics of phase ordering in phase 
transitions, is described, especially the emergence of condensates and 
scaling in Friedman-Robertson-Walker cosmologies, within a QFT framework.       
 
\subsection{Confinement: Supercharged Nucleus}

With the failure of $GUT$ theories to take care of the strong interaction 
sector $SU(3)$ of the Standard Model, the central issue of Confinement,   
which has had a long history of approaches ranging from the fundamental
to effective types, still remains an unsolved problem. There is a vast
literature on the subject, from Lattice QCD to various analytical methods for
non-perturbative QCD. Of these, 2 novel approaches to Confinement, which
are fairly self-contained, and stand out from the more conventional ones,
are included in Part A, leaving the rest for Part F. The first concerns  
an analogy to a super-charged nucleus, based on an old work of Pomeranchuk 
and Smorodinsky (1940), which offers the possibility of binding a particle 
in a small region of space. This method was elaborated in a set of THREE 
"Orsay Lectures"  by the late {\bf Vladimir Gribov} [6] during 1992-94. The 
basic idea is that if the charge $Z$ in a nucleus $N_Z$ is larger than a 
critical value $Z_c \approx 180$, then this nucleus will decay to an atom 
of charge $Z-1$ and a positron: $N_Z \rightarrow A_{Z-1} + e^+$. If the
product nucleus is unstable, the process gets repeated until the total
charge of the final product is so small that further decay is impossible.
Such a supercharged nucleus (a `resonance') cannot exist freely, but only 
inside an atom, hence is reminiscent of a `confined' state ! The region of 
stability of such a `superbound' atomic state, (mainly due to the Pauli 
principle), works out as $r_0 << r < 1/m$,  where $r_0$ is the radius of 
the nucleus, and $m$ the electron mass. In these three lectures, which 
are reproduced in this Book through the courtesy of his long term Associates
Dokhshitzer, Ewarz and Nyiri,  Gribov [6] gives a leisurely exposition of the 
detailed working of this mechanism on the confinement of heavy, followed 
by light, quarks. These ideas have since been extended by the Dokhshitzer
Group in their subsequent publications hep-ph/9807224  and hep-ph/9902279, 
but these are outside the scope of this Book.  

\subsection{Confinement: BRST Mechanism}     

The second approach concerns a perspective on confinement due to Nishijima 
who relates its mechanism to that of an unbroken non-abelian gauge symmetry 
in QCD. The logic of this method which was mostly pioneered by Nishijima, 
may be illustrated for the case of abelian QED as follows. Quantization of     
of the e.m. field requires "gauge-fixing", say by a covariant (Fermi)
gauge. This in turn requires introduction of the indefinite (Gupta-Bleuler) 
metric which, for the selection of physically observable states, must be 
eliminated by imposing the Lorentz condition on the state vector. There
are now 4 kinds of photons (2 transverse, 1 longitudinal, and 1 scalar), 
of which the two `scalar' photons must have negative norms, so as to ensure
manifest covariance of the quantization in the Minkowski space. 
\par
	Now to project out the physical subspace, one introduces a 
subsidiary (Lorentz) condition (a 4-divergence of a vector field) which
represents a {\it free, massless} field even under interactions. The photons
involved in this operator (called $a$-photons) are special combinations of     
longitudinal and scalar photons with {\it zero norm}. A second (orthogonal)
combination (called $b$-photons) also can be arranged to have zero norm. 
However the inner product of $a$- and $b$- photons is non-zero; they are
`metric partners' (somewhat akin to the 4-vectors $n_/mu$, ${\tilde n}_\mu$ 
defining a covariant null-plane: $n^2={\tilde n}^2=0$; $n.{\tilde n}=1$).
A physical state is defined as one that is annihilated by applying the
positive frequency part of the Lorentz condition. And since the S-matrix
in QED commutes with this 4-divergence, it transforms physical states 
into one another, without letting them out of this subspace which now
includes only $t$ (transverse) and $a$-photons, but $not$ $b$-photons.  
However the inner product of a physical state with  one $a$-photon, with
another physical state (with or without an $a$-photon), vanishes identically.
Thus $a$-photons give no contribution to observable quantities, and both
$a$- and $b$-photons escape detection ! This is called $confinement$ of                 
longitudinal and scalar photons in QED, a $kinematical$ phenomenon !
\par
	In QCD, on the other hand, not only $a$- and $b$-gluons, but also
the $t$-gluons are unobservable, giving a $dynamical$ orientation to the
confinement mechanism. While the basic logic and signature of confinement
for non-abelian QCD remains the same as above for abelian QED, some extra
ingredients of a highly technical nature are needed to bridge the gap. 
For not only the observable quantities now depend on the gauge parameter,
but the 4-divergence of the gauge field is no longer a free field ! To
eliminate the gauge-dependence of physical entities, Faddeev-Popov proposed
to average the path integral over the manifold of gauge transformations, 
resulting in a new term in the Lagrangian (Faddeev-Popop ghost), involving
a pair of anticommuting saclar fields whose violation of the Pauli theorem
on spin-statistics connection requires introduction of the indefinite 
metric, as in QED. However, the operator analog of the Lorentz condition 
is more tricky in this case. It is facilitated by a novel symmetry found
by Becchi-Rouet-Stora (BRS) which was originally used for renormalizing QCD.
Nishijima successfully exploited this symmetry to construct the requisite 
operator, and obtained a formal proof of confinement in the QCD case, as an
extension of the logic employed for QED. A qualitative sketch of this proof 
appears in the Article by {\bf {K Nishijima and M. chaichian}} [7].     
 
\section{Field Theory: Topological Aspects}

	An important sector of QFT that has come to occupy increasing
importance in the last two decades, concerns its Topological aspects, as 
a powerful tool to probe the geometry and topology in $low$ dimensions. 
This illustrates rather vividly the coming together of physicists and 
mathematicians, this time in building powerful links between quantum 
theory (through its path integral formulation) on the one hand, and the 
geometry and topology of low dimensional manifolds on the other. Indeed
it appears that the properties of low dimensional manifolds can be nicely
unravelled by relating them to infinite dimensional field manifolds, thus
providing a powerful tool for studying these manifolds.     
\par
	A unique characteristic of topological field theories is their
independence of the metric of curved manifolds on which they are defined. 
This makes the expectation value of the energy-momentum tensor vanish. 
Since the only degrees of freedom are topological, there are no $local$
propagating degrees of freedom. The operators are also metric independent. 
These features are addressed in some detail in a self-contained introductory 
Article by {\bf Romesh Kaul} [8] on topological QFT regarded as a meeting 
ground for physicists and mathematicians.

\subsection{CS Theory And Jones Polynomials}

Quantum $YM$ theories in $(2+1)D$ provide a field theoretic framework for 
the study of "knots and links" in a given 3-manifold, and illustrate the 
interplay of QFT and the topology of low dimensional manifolds. A 
striking result of this connection is that the famous "Jones Polynomials" 
of knot theory can be understood in 3D terms. This result was formally 
demonstrated by Edward Witten about a decade ago in a paper entitled
"Quantum Field Theory And The Jones Polynomial",  thus  fulfilling a 
long-cherished goal of an exact (non-perturbative) solution of a gauge 
field theory, for the first time in 3 dimensions. Witten showed that the
"Jones polynomial can be generalized from $S^3$ to arbitrary 3-manifolds,
giving invariants that are computable from a surgery presentation". Witten
further showed that these results shed new light on 2D conformal field
theory. In view of the historical importance of this pioneering work in the 
context of this Book theme, we reproduce (with permission from 
Springer-Verlag) the celebrated {\bf Witten paper} [9] (which had 
appeared in Commun.Math.Phys.{\bf 121} (1989) 351-399), in full.         

\subsection{Anomalies In QFT}

An interesting pathology of QFT which has rich topological overtones
is the problem of $anomalies$ which originated in the famous $ABJ$ (1969)
paper to resolve the problem of $\pi^0 \Rightarrow \gamma\gamma$ decay 
whose hitherto standard explanation in terms of partial conservation of 
axial current $(PCAC)$ used to fall far short of experiment.  The $ABJ$ 
paper finally resolved the issue by introducing an "anomalous" amplitude 
proportional to $F_\mu\nu {\tilde F}_\mu\nu$ in the $PCAC$ relation, whose 
interpretation brought into focus the pathology of $symmetry-breaking$ at 
the classical level through such "anomalies"  at the QFT level. Such 
`violation' of gauge symmetry through `anomalies' points to the need for 
their cancellation, which in turn constitutes an important constraint for 
physical gauge theories with $chiral$ coupling to fermions. In this respect, 
"global chiral anomalies" play a key role in the understanding of physical 
effects associated with topologically non-trivial gauge-field 
configurations, via the celebrated Atiyah-Singer Theorem. This subject 
is briefly reviewed by {\bf Haridas Banerjee} [10] in this Book.        

\subsection{Coherent States In QFT} 

Still another sector of QFT with topological (geometric) features,
is the subject of {\it Coherent States} which has grown rapidly since 
its birth 36 years ago at the hands of Glauber and Sudarshan
[R.J.Glauber, Phys.Rev.{\bf 130}, 2529 (1963); E.C.G.Sudarshan, Phys.Rev.
Lett.{\bf 10}, 277 (1963)], although the basic idea dates back to the 
founder of Quantum Mechanics himself [Erwin Schroedinger:Naturwissenshaften,
{\bf 14}, 644 (1926)] in connection with the quantum states of a harmonic 
oscillator, i.e., almost immediately after the birth of quantum mechanics. 
Coherent States have 3 main properties: coherence, overcompleteness and 
intrinsic geometrization, all of which play a fundamental role in QFT. These
include the calculation of physical processes involving infinite number of
virtual particles; the derivation of functional integrals and various
effective field theories; and last not least,  the exploration of the 
origins of topologically non-trivial gauge fields and the associated 
(gauge) degrees of freedom. All these topics are addressed systematically 
in a perspective, self-contained review by {\bf Wei-Min Zhang} [11].  

\subsection{Pancharatnam-Bargmann-Berry Phase}
        		
An outstanding example of a topological aspect in quantum mechanics (which 
may be termed `field theory with a finite number of degrees of freedom'), 
is provided by the existence of a "geometric phase" in quantum theory which 
had remained obscured from public view until rather recently when M.Berry 
(1984) drew attention to it under the term "quantum adiabatic anholonomy".
Historically, however, the existence of this pathology in physics had 
first been noted by S.Pancharatnam (1956) in the regime of classical
polarization optics, but this important work had somehow gone by default.
A similar fate befell a second attempt by V.Bargmann (1964) to resurrect
this idea in the context of Wigner's theorem on the representation of
symmetry operators in quantum mechanics. It was only after the work of Berry
that its full implications were appreciated within the physics community,
but its connection with the Pancharatnam and Bargmann phases was left
unattended. In an instructive Article, {\bf N Mukunda} [12] describes 
these developments in a proper perspective by emphasizing the mutual
connections among these ideas. He also describes the subsequent developments
to date, by relating these phases to the presence of a complex vector space 
and the effect of group action among them. He then goes on to show that the
geometric phase is the simplest invariant expression under certain groups
of transformation acting on curves in Hilbert space.   

\subsection{Skyrmion Model for Confinement}

A confinement mechanism with topological overtones is offered by the large 
$N_c$ limit of QCD which has played a crucial role in unifying its premises  
with a solitonic, hadron-based approach that is known as the $Skyrme model$
which was discovered by Skyrme (1961), just before quarks (1964) were born.
Skyrme's novelty was to provide a model in which the fundamental fields
consisted only of {\it pions}, wherein the nucleon was obtained as a 
certain classical configuration of pion fields. The apparent contradiction 
of making Fermi fields out of Bose fields was resolved by demanding
a non-zero "winding number" for this (classical) field configuration,
thus giving the "Skyrmion" the status of a topological soliton, which is
a solution of a classical field equation with localized energy density.
\par
	On the face of it, the Skyrme scenario looked so different from 
the conventional picture of nucleons as a `white' composite of 3 `colored' 	
quarks bound together by their interactions with $U(3)$ gauge fields, that
a reconciliation between the two pictures appeared rather remote. It 
turned out however that the Skyrme model could be a plausible approximation
to the orthodox QCD picture, one in which a key role is played by the
large $N_c$ limit of the latter. The logic goes roughly as follows.
\par
	Despite the increasing strength of QCD at low energies, it is
plausible that the pseudoscalar mesons as $q{\bar q}$ {\it composites},
could still interact relatively weakly with each other, thus permitting
the formulation of some {\it effective} Lagrangian for the pions, subject
of course to the correct symmetries of the underlying gauge theory, which
includes a (spontaneously broken) chiral $SU(N_f) \otimes SU(N_f)$ 
flavour $(N_f)$ symmetry that  allows `massless' pseudoscalars to
co-exist with massive scalars. An effective Lagrangian on these lines 
may be obtained from "a non-linear realization of chiral symmetry", 
without the explicit appearance of scalars, a structure which has an
uncanny resemblance to the very Lagrangian obtained by Skyrme (1961).
\par
	How about the baryons in this QCD-motivated "chiral perturbation 
theory" picture ? It is here that t'Hooft's (1974) large $N_c$ limit comes 
into play, with the proportionality to $N_c$ for the baryon mass being the
signal that the baryon state under study is a soliton of the effective
meson theory initiated by Skyrme. In a perspective review of the Skyrme 
model approach, {\bf {Joseph Schechter and Herbert Weigel}} [13] trace 
its connection with QCD in the large $N_c$ limit,  and discuss the 
properties of light baryons treated as solitons, within the framework of 
an effective Lagrangian of QCD containing only meson degrees of freedom.              

\section{Formal Methods In QFT: Selected Topics}

	The universal language of QFT and its powerful techniques broke
fresh ground through the establishment of the equivalence of its tenets 
with those of Statistical Mechanics which had traditionally been developed 
on entirely `classical' lines. In the words of A.M.Tsvelik (QFT in CMP,
Camb.Univ Press 1995), this equivalence may be succinctly expressed 
by the following statement: " QFT of a $D$-dimensional system can be 
formulated as a statistical mechanics of a $(D+1)$-dimensional system. 
This equivalence .... allows one to get rid of non-commuting operators
and to forget about time ordering, which seem to be the characteristic
properties of quantum mechanics....". The Path Integral formulation of 
QFT which is the key element in dispensing with the problem of non-commuting
operators in QFT, has had a crucial role in bringing about this vital 
correspondence of QFT with the partition function in quantum statistical 
mechanics (QSM). Armed with the powerful techniques of Renormalization 
Group Theory (RGT), this new approach has opened up a whole vista of 
applications to new emerging areas like critical phenomena in condensed 
matter physics. 

\subsection{Unified View of QFT and QSM}
 
	An important outcome of a unified view of QFT and Quantum 
Statistical Mechanics has been the emergence of two new areas: Euclidean 
Field Theory, and Finite Temperature Field Theory. Actually the origins of 
the former date back to the Fifties at the hands of Wick ("Wick rotation" 
for the Bethe-Salpeter equation) and Schwinger (as a possible direction for 
the evolution of QFT), wherein the transition from Minkowski to Euclidean 
space (via analytic continuation from real to imaginary "time") was perceived 
as a means of curing many ills in QFT, such as positivity and finiteness of 
norms in the computation of physical quantities. In more recent times, the 
Euclidean formulation of QFT has led to an interesting relationship between 
"stochastic mechanics" (Nelson) and the Feynman-Kac formulae for Green's 
functions expressed as path integrals. In a crisp Article in this Book, 
{\bf R.Ramanathan} [14] provides a formulation of QFT in Euclidean space-time,
to bring out the basic ideas of the Euclidean formulation, as well as the
above relationship between the Nelson and Feynman-Kac formulations.   
\par
	Finite Temperature Field Theory on the other hand, (in contrast to 
zero temperature for Euclidean QFT), provides access to a much wider class 
of complicated quantum mechanical systems, and addresses questions like 
thermal averages in QFT, symmetry restoration in theories with spontaneous 
symmetry breaking, and indeed the evolution of the universe at early times 
(from the high temperature phase). More recently, chiral symmetry-breaking 
phase transitions, especially the "confinement-deconfinement" phase 
transitions in QCD leading to quark-gluon plasmas $(QGP)$, have acquired 
great interest in view of planned experiments on heavy ion collisions to 
detect $(QGP)$. A few selected topics in Finite Temperature Field Theory 
are treated in an informative Article by {\bf Ashoke Das} [15] in this Book.            

\subsection{Integrable Systems: Toda FT}

Although most approaches to QFT have been traditionally associated
with  linear  partial differential equations, (e.g., Schroedinger, 
Klein-Gordon, Dirac, Proca), {\it non-linear} equations, (i.e., equations
where the potential term is non-linear in the field $\phi$), have also been
known for some time. Among the earliest non-linear wave equations known in 
physics are the $Liouville$ and $Sine-Gordon$ equations. The Liouville 
equation in 2D arose in the context of a search for a manifold with
constant curvature, something like covering the surface with a fishing net 
whose arc length is constant (knots do not move!), while the `threads' in the 
net correspond to a local coordinate system on the surface. The "field"
$\phi$ in the Liouville equation is the phase space density $\rho$ 
satisfying the equation $\partial_x\partial_y\rho=\exp\rho$, where $x,y$ 
are the local orthogonal coordinates. The Sine-Gordon $(SG)$ equation 
has a similar structure, with $\exp$ replaced by $\sin$ on the RHS.      
Variants of these equations, e.g., adding a `mass' term $m^2\phi$ on the 
LHS, and/or the hyperbolic replacement of $\sin$ by $\sinh$, etc,
give rise to several more varieties of similar types.  A third type of
non-linear equation which has received much attention, is the so-called
$KdV$ equation $u_t-6uu_x+u_{xxx}=0$, with interesting properties like an
infinite number of conservation laws. The corresponding conserved quantities
can be used as Hamiltonians for an integrable system ($KdV$ hierarchy).          
A striking feature of such non-linear equations is an infinite number of 
conserved quantities, which imply that the solutions of these systems must 
be infinitely restricted. This results in such solutions being quite stable
structures ($solitons$) which retain their shapes even after collisions.
\par
	An interesting class of coupled non-linear equations was introduced 
by M.Toda (1967) to describe a 1D crystal with non-linear coupling between 
nearest neighbour atoms. These (lattice) models also admit $soliton$ 
solutions which reduce to the $KdV$ equation in the continuum limit. At
the `field' level, such models (with exponential `potentials') simulate 
a general class of non-linear equations--called Toda Field Theory--which
include the Liouville and Sine-Gordon equations as special cases. For the 
solution of these equations, a general method of "inverse scattering" was
proposed by Gelfand-Levitan. The logic of this method is to convert, via
a suitable transformation, the original non-linear equation to an
equivalent $linear$ equation, and study the evolution of the latter, 
more or less according to  standard methods already developed for them 
(including group-theoretic, Lie-algebraic, etc methods). The inverse 
scattering method paved the way to connections with other known models
of QFT, such as conformally invariant FT and the Hamiltonian reduction of
Wess-Zumino-Witten model. Similarly the $KdV$ equation is related to the
4D Yang-mills theories, thus providing a connection of the latter with 2D 
integrable models. In an instructive, self-contained article on this subject, 
{\bf Bani Sodermark} [16] gives a perspective view of integrable systems
with special reference to the Toda Lattice hierarchy, and reveals the
connections of such non-linear field theories with other sectors of QFT. 

\subsection{Light-Front Dynamics}

Dirac laid the foundations of QFT, not only through his famous Equation,
but at least with 2 more seminal contributions within a year's gap from
each other: a) light-front (LF) quantization [Rev.Mod.Phys.{\bf 21}, 392 
(1949)]; b) constrained dynamics [Can.J.Math.{\bf 2}, 129 (1950)]. In the
former, he suggested that a relativistically invariant Hamiltonian theory
can be based on different classes of initial surfaces: instant form 
($x_0=const$); light-front (LF) form ($x_0+x_3=0$); hyperboloid form 
($x^2+a^2 <0$). The structure of the theory is strongly dependent on these 
3 surface forms. In particular, the " LF form" remains invariant under $7$ 
generators of the Poincare' group, while the other two are invariant only 
under $6$ of them. Thus the LF form has the maximum number ($7$) of 
"kinematical" generators (their representations are independent of the 
dynamics of the system), leaving only 3 "hamiltonians" for the dynamics. 
\par
	Dirac's  LF dynamics got a boost after Weinberg's discovery of the 
$P_z =\inf$ frame which greatly simplified the structure of current algebra.
The Bjorken scaling in deep inelastic scattering, supported by Feynman's 
parton picture, brought out the equivalence of LF dynamics with the 
$P_z=\inf$ frame. The LF language was developed systematically within the QFT
framework by Kogut-Soper (1970), Leutwyler-Stern (1978), Srivastava (1998)
and others. The time ordering in LF-QFT is in the variable $\tau=x_0+x_3$, 
instead of $t=x_0$ in the instant form. And despite certain technicalities, 
the LF dynamics often turns out to be simpler and more transparent than the 
instant form, without giving up on the net physical content.  This is borne
out from comparative studies: of spontaneous symmetry breaking on the LF ;
of degenerate vacuum in certain $(1+1)D$ QFT which are exactly soluble and 
renormalizable (e.g., the Schwinger model and its chiral version); of 
chiral boson theories; and of QCD in covariant gauges. Indeed, the LF 
quantization of QCD in the Hamiltonian form bids fair to be a viable
alternative to the lattice gauge theory for calculating non-perturbative
quantities. Removal of constraints by the Dirac method gives fewer 
independent dynamical variables in the LF formalism than in the instant
form; for this reason, LF variables have found applications even in String 
and $M$-theories. In an instructive self-contained review (with a rich
collection of references), {\bf Prem Srivastava}[17] gives a detailed review 
of most of these topics in a leisurely and systematic manner, and leads 
the interested reader all the way to the frontier with several new results. 

\subsubsection{$2D$ Field Theory}

$2D$ models in QFT have also been of great interest in the contemporary
literature. Such theories reveal some remarkable features, such as 
fermion-boson equivalence, which facilitates the solution of fermion-FT
in terms of its bosonized version. This concept of bosonization in turn has 
been useful in the understanding of $4D$ phenomena that can be described
by an effective $2D$ FT, such as the demonstration of  quark confinement 
in exactly soluble $2D$ models [Casher-Kogut-Susskind (1973)]. Another
important discovery in $2D$ FT concerns an "anomaly-generated" mass 
[Jackiw-Rajaraman (1985)] for the gauge boson in the Chiral Schwinger
model. (This mechanism may be contrasted to the standard Higgs mechanism for 
generating the vector boson mass via spontaneous symmetry breaking). The
"anomaly" here stands for the loss of the conservation property due to
quantum corrections involved in the quantization of the gauge theory. This
disease in turn needs Dirac's second weapon for cure: Constrained dynamics.
In a short perspective article in this Book, {\bf Dayashankar Kulshreshtha} 
[18] reviews the constrained dynamics and local gauge invariance of several 
$2D$ FT models, in both Instant and LF forms, and in so doing, brings out 
the detailed working of the BRST formalism as applied to such $2D$ models. 

\subsection{Constrained Dynamics}  
	
To recall the essential elements of a constrained dynamical system, which 
includes most systems of physical interest (e.g., QED, QCD, Electroweak and 
Gravity theories), it is characterized by an $over-determined$ set of
coordinates. These are best kept track of within a Hamiltonian formulation,
which has a natural place for all the coordinates (canonical and redundant),
so that the complete set of constraints emerges easily. The nature of these
constraints in turn is determined by the structure of the matrix of Poisson
brackets (PB) of the constraints of the theory, which also carries the 
signature of whether or not the underlying theory is gauge invariant (GI). 
Thus if this PB matrix is singular, then the set of constraints is $first
class$, and the theory is GI. On the other hand, if this matrix is non-
singular, then the set of constraints of the theory is $second class$, 
and the theory is non-GI. (Indeed this is often taken as a criterion for
distinguishing a GI from a non-GI system). These GI systems are then 
quantized under some appropriate gauge choices, or "gauge fixing" (GF).
Now in the usual Hamiltonian formulations of a GI theory under some GF's,
one necessarily destroys the gauge invariance, since the GF corresponds to
converting the first class constraints to second class constraints. To
quantize a GI theory by maintaining gauge invariance despite GF, one needs
the more general BRST (1974) formulation, wherein the theory is rewritten
as a quantum system with generalized GI, called BRST invariance. This in
turn requires enlarging the Hilbert space, and replacing the gauge 
transformation by a BRST transformation which involves the introduction
of (anti-commuting) Faddeev-Popov $ghost fields$. This amounts to embedding
the GI system into a BRST invariant system (but isomorphic to the former),
whose unitarity is guaranteed by the conservation and nilpotency of the
BRST charge.   
\par       
	Thus the  Dirac[Can.J.Math.{\bf 2}, 129 (1950)]-Bergmann 
[Phys.Rev.{\bf 83}, 1018 (1951)] theory of Constraints lies at the root of
(Hamiltonian) description of interactions in QFT  based on Action principles 
which, due to the requirements of Lorentz, local gauge, (and/or 
diffeomorphism) invariances, must employ singular Lagrangians. This is
generally adequate for the study of simple gauge theories (controlled by 
some Lie groups acting on some internal spacein Minkowski space-time), via 
the covariant approach based on BRST symmetry which, at least for 
infinitesimal gauge transformations, allows a regularization and 
renormalization of the relevant theories within the local QFT framework.
On the other hand, the gauge freedom of theories that are invariant under 
diffeomorphism groups of the underlying space-time (e.g., in general 
relativity or string theory) is encumbered by the arbitrariness for the 
observer in the "definitory properties" of space-time and/or the measuring 
apparatus;[see {\bf L.Lusanna}-this Book]. Such ambiguities affect bigger 
issues like: the understanding of $finite$ gauge transformations; the 
Gribov ambiguity in the choice of function space for the fields; proper 
definition of relativistic bound states vis-a-vis quark confinement; and 
last not least the conceptual and practical problems posed by gravity. 
These require a fresh look at the foundations of QFT to know if we: 
i) understand the physical degrees of freedom hidden behind gauge and/or 
general covariance; ii) can meanfully reformulate the physics (both
classical and quantum) in terms of them. Logically this would amount to
abandoning local QFT for non-perturbative interactions, and a reformulation 
of relativistic theories to allow natural coupling to Gravity. These and
allied issues are addressed in a state of the art review by 
{\bf Luca Lusanna} [19], aimed at a unified reformulation of the 4 basic 
interactions in terms of Dirac-Bergmann observables, with emphasis on the 
open problems--mathematical, physical and interpretational. 
        
\section{Extension Of QFT Frontiers}  

A long term ambition of QFT has been the dream of unificication of all the 
gauge fields with the Gravitation Field whose quantization has all along
posed a big challenge in its own right. [A major difficulty in the way of 
unification of this sector with the other three, as was once succinctly
put by Abdus Salam, lies in the "spin mismatch" of their respective fields
(vector vs tensor), which would militate against a common strategy].  
Nevertheless such a unification was to come about from an entirely new
paradigm which envisaged extension of the original tenets of Field Theory  
based on a point particle description to one with $Strings$. In this
Section we offer a panoramic view of some major theoretical developments 
from seemingly unrelated angles, which, apart from their impact on Physics
in their own right, have provided some key ingredients converging towards 
the emergence of modern $String Theory$.  These developments which may be 
termed $Supersymmetry$ (SUSY), $Conformal Field Theory$ (CFT), and 
$Duality$, are outlined next. 

\subsection{SUSY In Field Theory}
 
In its march towards Unification, Field Theory has continued to break new 
ground in several directions. An important step in Unification was marked 
by the discovery of Supersymmetry $(SUSY)$, introduced in the early seventies 
by a galaxy of authors in the context of 2D QFT (Gervais-Sakata) as well as
in 4D QFT (Golfand, Likhtman, Akulov, Volkov, Wess and Zumino), for a
unified understanding of the two known forms of matter--bosons (integral
spins) and fermions(half integral spins)--hitherto regarded as two distinct 
field types, with commuting and anticommuting properties respectively. The
new symmetry between bosons and fermions may be incorporated within the
definition of a single "Superfield", with transformations inter-relating 
the two constituents, so that $SUSY$ becomes a part of space-time symmetry
implied by relativistic invariance. The Gauge principle too admits of a
corresponding extension to unify both these sectors.
\par
	What are the motivations for such a lavish extension of space-time
symmetry ? Apart from its aesthetic appeal, there are some theoretical
considerations of a more concrete nature which are dwelt on in this Book
through two complementary reviews of $SUSY$ in Field Theory, (with 
special reference to Particle Physics), by two leading experts in the field: 
{\bf Rabi Mohapatra} [20] and {\bf Norisuke Sakai} [21] respectively. 
According to Sakai [21], the most important motivation for $SUSY$ is the 
$Gauge hierarchy problem$ showing up via the vastly different mass scales 
of the electroweak ($M_W$) vs the "$GUT$-theoretic" ($M_G$): $M_W^2/M_G^2$
$\approx 10^{-28}$. A similar gap exists between the "GUT" vs Planck 
(gravity) mass scales: $M_G^2/M_{pl}^2$$ \approx 10^{-6}$. 
\par
	To account for this phenomenon, it is necessary to invoke a suitable 
$Symmetry$ reason which may be precisely formulated by the so-called  
"naturalness" hypothesis (t'Hooft 1979) which demands that a system acquires
a higher symmetry as a certain (small) parameter goes to zero, e.g., chiral
symmetry occurs when a (small) fermion mass goes to zero; or a local gauge
symmetry corresponds to the vanishing of a vector boson mass. Now the mass 
scale $M_W$ of weak bosons arises from the vacuum expectation value 
$<\phi>_0 \equiv v \neq 0$, related to the mass $M_H$ of the Higgs scalar
field $\phi$. So to regard the gauge hierarchy problem as the result of some
symmetry breaking, we must give a $Symmetry$ reason to make the Higgs scalar
mass vanishingly small. Classically a vanishing scalar mass corresponds to
a symmetry called scale invariance, which however cannot be maintained
quantum mechanically. In a perspective review on this subject, 
{\bf Norisuke Sakai} [21] argues for "Supersymmetry" between the Higgs scalar 
and a spinor partner as a good option: Chiral symmetry gives zero mass to 
the latter, while $SUSY$ makes the former massless (through a cancellation 
of the respective contributions to the self-energy loops).   
\par	 
	In a complementary perspective review on the same subject, 
{\bf Rabi Mohapatra} [20] stresses the versatility of $SUSY$ as a tool for 
understanding many unsolved problems of physics: a) improvement in the 
singularity structure of local fields for understanding the disparate 
scales of Nature (e.g., Electroweak vs Gravity); b) possibility of unifying 
Gravity with the other forces by making $SUSY$ local instead of global; 
c) prospects of understanding $non-perturbative$ properties of field 
theories, hitherto considered `impossible' in non-$SUSY$ form.        
\par
	As to the manifestations of $SUSY$ in a real world, this "Bose-Fermi" 
symmetry is supposed to be badly broken, so that any search for superpartners
(bosons vs `bosinos'; fermions vs `s-fermions') has so far yielded zero 
dividends. On the other hand the formulation of Supersymmetry in 
non-relativistic quantum mechanics is relatively free from constraints. 
Indeed, since Schroedinger (1940) noticed the existence of well-defined 
"supersymmetric partners" for the energy levels of a given quantum 
mechanical system, many applications to such systems (including nuclear and 
condensed matter physics), have kept pace with the rapid strides of SUSY 
in field theory in recent years. Indeed, the existence of SUSY partners 
in the energy levels of (appropriately chosen) even vs odd nuclei have 
been systematically established by group theoretic methods (interacting 
boson models, etc). Similarly, in solid state physics, an interesting 
correspondence has been observed between the critical behaviour of a 
`spin system' in random magnetic fields in $d$ dimensions, and that of 
the spin system without the random magnetic field in $d-2$ dimensions. 
This "dimensional reduction" may be traced to an underlying $SYSY$ for
the spin system in random magnetic fields (see {\bf N Sakai}--this Book).   
\par    
	In the absence of discovery of $SUSY$ partners in Field Theory, 
the benefits from $SYSY$ have so far been purely theoretical, varying from 
reduction of the degrees of divergence arising from various loop integrals 
in standard field theory (by at least two orders), to a heavy reduction in 
the number of dimensions (from 26 to 10) needed for  self-consistency in a 
string theoretic formulation. The Articles by Mohapatra [20] and 
Sakai [21] between them provide quite a complementary description of the 
$SUSY$ formalism in QFT, together with an glimpse of the recent developments. 
And apart from its applications in particle physics, this formalism also 
serves as a background to the vast field of supersymmetric string theory. 

\subsection{CFT}
          
	An independent insight into the origin of String Theory comes from 
the role of $Conformal$ $Field$ $Theory$ (CFT), viz., conformally invariant 
QFT in 2D(imensions), not only as as a vital ingredient of its anatomy, but
also with firm hold on other disciplines like condensed matter physics. 
The CFT route to the evolution of String Theory is sketched in this Book as 
part of a bigger (historical) survey by {\bf Werner Nahm} [22], tracing a 
whole sequence of developments in QFT right from its (Dirac) beginning, and 
encompassing in the process several other areas of physics on which CFT 
has had a decisive impact. In this saga, the interplay of physical intuition 
and mathematical rigour has brought together the practitioners of these 
respective disciplines, though not necessarily working in tandem. On the 
one hand, the beauty and transparency of CFT have made for a rich variety of
intellectual exercises in abstract mathematics (with new emerging areas 
like automorphic groups, K\"ahler-Einstein metrics, etc), and on the other,
facilitated the study of intensely practical physical systems such as 
continuous phase transitions in condensed matter physics. 
\par
	The impact of CFT on string theory has had its origin in several 
theoretical developments: the Thirring model in 2D; Skyrme's idea of 
the equivalence between Fermions and Bosons; Coleman's equivalence 
theorem on the Thirring Model versus the Sine-Gordon equation 
(despite their apparent dissimilarity); and the role of conformal 
invariance in the structure of Wilson's Renormalization Group equations.
To recall the essentials of Conformal invariance, this symmetry is  
satisfied in the absence of any `scale' dimension. Examples are Maxwell's
Equations in free space; Dirac equation for massless fermions which satisfy
conformal invariance. The 2D Thirring model, which may be regarded as a
basic ingredient of string theory, also has this property due to absence
of a scale dimension. Using this mathematical picture, the string may be
regarded as a 1D object in space spanning a world sheet (a Riemann surface)
embedded in 2D space-time, where a point on the string is represented by 
$X^\mu(\sigma, \tau)$; $\sigma$, $\tau$ being the 2 world sheet coordinates.
\par
	The impact of CFT has been no less impressive in the domain
of condensed matter physics (CMP) where there exist a rich class of QFT's 
exhibiting the structure of conformally invariant fields, such as in 2D 
surface coatings.  Thus at a critical temperature $(T_c)$, the long range 
fluctuations of arbitrary scales make irrelevant the details of molecular 
structure, and the theory approaches a $continuum limit$, with no
visible scale dimension to keep track of. Indeed in this limit, the 
correlation functions behave like the Euclidean $n$-point functions of
standard QFT with conformal invariance properties. Nahm [22] discusses an
interesting correspondence between the Ising model in CMP and Thirring  
model in QFT. The equation satisfied by the spin waves of the Ising model
is formally identical to the 2D Dirac equation for massless fermions.    
Indeed condensed matter physics provides a more stable and economical
background for testing these ideas than the expensive HEP laboratories !

\subsection{String Theory Via Duality}

Perhaps the most startling "revolution" in Physics to date which had its 
origin in QFT, has been the String Theory, and its successive "Avatars"
(incarnations), aimed at unifying all the forces of Nature (from orthodox 
gauge theories of strong, e.m. and weak interactions, all the way to
gravity). An orthodox route to its evolution may be attributed to the 
strong interaction problem in QFT, which has had wide ramifications from 
vastly different angles, each providing an independent insight into its
mysteries. A very promising approach to strong interactions came from the 
$Duality$ Principle which has had a long history (prehaps traceable to
the Bootstrap hypothesis), based on the equivalence of the direct channel
(resonances) and crossed channel Regge poles with a universal slope
$\alpha'\approx 1 GeV^{-2}$. An explicit realization of this idea was 
achieved via the Veneziano representation for $4$-point amplitudes 
satisfying the requirements of duality and crossing symmetry, which was
soon generalized to $N$-point amplitudes satisfying the same properties.
Through a path integral representation of such amplitudes,  Nambu,
Nielsen and Susskind recognized that these amplitudes describe a 1D 
(string-like) object moving in space,  with the inverse of the universal 
Regge slope identified as the "string" tension $T$. The "string"
interpretation was further reinforced by a subsequent representation due to
Virasoro, with very similar properties. And its promise of relevance to 
particle physics (despite stiff competition from QCD!) got a boost from the 
Scherk-Schwarz (1974) observation that such a "string theory" could serve as
a candidate for incorporating gravity in its ambit, on the ground that the 
massless spin-2 particle appears naturally in the closed string spectrum. 
To that end the string tension $T$ needed to be increased by $19$ orders of 
magnitude (up to the Planck scale!) to qualify for a viable theory of gravity.
The conceptual gap was finally bridged by the seminal work of Green-Schwarz 
(1984) who succeeded in constructing a consistent $10D$ super Yang-Mills 
theory coupled to supergravity which is free from anomalies only for 
certain gauge groups ($SO(32)$ or $E_8 \times E_8$). This work, perhaps for 
the first time, showed real prospects for unifying the fundamental forces. 
\par
	The String Theory has grown by leaps and bounds during the past
decade, and its vast ramifications have grown to such formidable literature  
over the past decade, that a minimal justice to it would itself require 
several volumes of review. Nevertheless, after a short overview by the Master,
{\bf John Schwarz} [23] of the subject, a panoramic account of the major 
developments in this exciting field (together with an exhaustive set of 
references) is given in a perspective Article by {\bf Jnanadeva Maharana} 
[24]. Schwarz [23] views the different superstring theories (and an extension 
called $M-theory$) as different facets of a unique underlying theory going 
beyond ordinary QFT's. However, recent duality conjectures suggest that a 
more complete definition of these theories may come from the large $N$ limits
of suitably chosen $U(N)$ gauge theories; (see {\bf L Bonora} [25] below). 
The Maharana Article [24] leads the interested reader through several 
stages of its development, from i) perturbative aspects of $ST$; 
successively through ii) $Duality Symmetries$ as a characteristic of String 
Theories (ST); iii) $M$-theory as a unified view of the $five$ 
perturbatively consistent $ST's$; iv) microscopic understanding of Black 
Holes, and so on, all the way to the frontiers of the field.   
\par
	Attempting to cover the later stages of development in this 
rapidly  growing field, {\bf Loriano Bonora} [25] reviews some advances in
the study of the relation between Yang-Mills $(YM)$ theory and strings,
based on the classical $YM$-theory solutions ({\it Riemannian instantons})
which are 2D solutions describing Riemann surfaces in the strong coupling
limit. Strictly, such relations historically date back all the way to
't Hooft ('74) through his famous $1/N_c$ expansion for large $N_c$, wherein 
the dominant Feynman amplitudes correspond to the 2D Riemann surfaces. 
This `natural connection' with strings was subsequently upgraded to a 
concrete shape via studies of 2D QCD (for string-like properties), which 
was further generalized to a connection between conformal super-$YM$ and
super-string theory of type $IIB$, in the large $N_c$ limit. The {\bf Bonora}
Article reveals, among other things, a direct link between String Theory
and non-abelian $YM$ theory, through the emergence in the latter of 
classical solutions modelled over Riemann surfaces, leading to a "string"
interpretation. Historically, this came about only after the proposal of
the $Matrix Theory$, which in the large $N_c$ limit converges to the 
(non-perturbative) $M-Theory$.    
      
\section{CS Field Theory And Condensed Matter Physics}
 
While the dominant concern in Field Theory has been in the 
traditional domain of particle physics, its powerful language and 
tecnniques have found profitable employment over a much wider domain,
which comprises topics in Condensed Matter Physics, and newly emerging
fields like Quantum Hall Effect, fractional statistics and $Anyons$.
These phenomena lend themselves to QFT treatment in $(2+1)$ dimensions,
where the celebrated "Chern-Simon" $(CS)$ term plays a key role (see
also Part B on Topological Field Theories). 
\par
	What are the special features of QFT in $(2+1)D$, and what specific 
role does the $CS$ term play in this reduced space-time continuum ? Perhaps
the most striking feature is the appearance of $fractional statistics$ !
For, whereas in 3 (or higher) space dimensions, all particles must either 
be bosons (integral spin) or fermions (half-integral spin), in 2 space 
dimensions, the particles can have any fractional spin/statistics with
impunity ! Such particles are called $Anyons$. Now since the usual 
spin-statistics relation follows from the premises of the standard 4D
relativistic QFT, it is natural to ask if $Anyons$ can be understood
from the corresponding 3D QFT. The question goes beyond mere academic 
interest since lower dimensions can be effectively realized in the 
physical world through the "freezing" out of certain degrees of freedom,
(e.g., in a strongly confined potential, or at low enough temperatures),
so that these `quasi-particles' may well exhibit anyon-like properties. 
And indeed experiments on Quantum Hall Effect (QHE) have revealed the 
existence of fractionally charged excitations (thus implying anyons). 
\par
	A critical discussion on the question of anyons and fractional
statistics in $(2+1)$ dimensions, with particular reference to the role of
the Chern-Simons $(CS)$ term in 3D QFT, is given by {\bf Avinash Khare} 
[26] in a perspective Article on the subject in this Book. To that end,
Khare clarifies the definition of "quantum statistics" which relates to
the "phase" picked up by a wave function when two identical particles are
$adiabatically exchanged$, as distinct from the usual definition of
permutation symmetry for two identical particles. [While both definitions
coincide for 3 and higher dimensions, they differ in 2 dimensions]. He then
discusses in detail the main properties of the $CS$ term, especially its
role as a gauge field mass term, in whose presence anyons can appear in one 
of two different ways: i) as a soliton of the corresponding QFT ; or ii) as  
fundamental quanta carrying fractional statistics. So far, the state of the
art is based on non-relativistic QFT, wherein the $CS$ term provides an 
effective cushion against a non-local formulation of anyon fields, thus 
facilitating a `local' formulation. However a full-fledged relativistic
QFT formulation is not yet feasible.     
\par
	Perhaps the most tangible success from $CS$ fields so far is a 
natural understanding of the Quantum Hall (QH) Effect. A state-of-the-art 
review by {\bf R Rajaraman} [27] puts this topical subject in perspective. 
We summarize some essential features of a QH system, from his own account. 
A QH system which is defined as "quasi 2D layers of electrons trapped in the
interface of semi-conductors, at very high magnetic fields and very low
temperatures, has revealed many remarkable features ". Particularly 
interesting is the presence of certain states characterized by the 
so-called "filling fractions" $(\nu)$ which are either integers, or
certain $odd denominator$ fractions; $\nu= hc{\bar \rho}/{eB}$, where
${\bar \rho}$ is the mean electron density, and $B$ the applied field.
The special states corresponding to these $\nu$-values show extremely flat
plateaus in Hall conductivity which (in units of $e^2/h$) are exactly equal
these values to within an accuracy of $1$ in $10^7$ ! These features are
very universal inasmuch as the details of the material seem irrelevant. 
It was earlier recognized that the electrons in these QH states form an 
incompressible fluid, described by "Laughlin wave functions" (which are
reminiscent of Jastrow-type correlations in nuclear wave functions). A
more analytical study of these empirical functions suggested a 
Landau-Ginsberg type scenario for the QHE in terms of an order parameter
field (subsequently to be identified with a Chern-Simons field), thus 
formally bringing this subject within a 3D QFT network.
\par
	The analogy of the order parameter field in QHE to that obtaining
in superconductivity of the Landau-Ginsberg description, is of course not 
a literal one since there are no bosonic Cooper pairs in QHE. Indeed 
in this 3D QFT scenario, the "anyons" (Chern Simons fields) have an 
intermediate status between bosons and fermions. However for the special 
case when the anyon angle is an odd multiple of $\pi$, a composite of the 
electron with an $odd$ number of flux tubes, effectively amounts to 
constructing a "bosonic" analogue of Cooper pairs from out of "fermions"   
which now provides the desired order-parameter (CS) field  operating in 
the plateau of the QH system. Rajaraman reviews a formal QFT procedure
for constructing such CS gauge fields, as well as the formulation of their
dynamics at the 3D QFT level. As to the connection of the CS gauge fields
at the first quantized level, these are of course expressible in terms of 
the "phase angles" involved in the exchange of electrons in an $N$-
electron wave function in 2D (see also  Khare [26] in this Book).           	          

\section{QCD-Motivated Strategies For Strong Interactions}

   Turning now to the strong interaction problem in the standard field 
theoretic picture, its prime candidate, QCD, has since its birth been
beset with problems of reliable calculational techniques to deliver results.
An introductory overview of several approaches [symmetries, effective 
Lagrangians and Wilson expansions] to deduce hadron properties from QCD 
is sketched in the Article by {\bf Olivier Pene} [28], aimed at establishing 
a link between perturbative and non-perturbative QCD via lattice methods.
We now go into more specific details of a few principal QCD-based methods.  

\subsection{QCD Sum Rules}

 To recall the main signatures of the prime candidate, QCD,  which 
it shares with any non-abelian gauge theory, are expressed by a two-fold 
pattern: i) decreasing coupling strength at shorter distances 
(Asymptotic Freedom); and ii) increasing coupling strength at longer 
distances ($confinement$). The former is fairly well understood, and 
provides a perturbative basis for calculating QCD effects in high energy 
processes. In particular, the powerful method of "QCD Sum Rules", based on 
Wilson's Operator Product Expansion $(OPE)$, was developed by Shifman-
Vanstein-Zakharov for the study of non-perturbative QCD in a large variety  
of applications from hadronic masses (with two-point functions), coupling 
constants, form factors (with three-point functions), and reactions (four
point functions). The basic philosophy is one of a $duality$ between two 
ways of representing a correlator: i) $OPE$ with various "twist" terms 
(vacuum condensates, treated as free parameters of the theory) representing 
successive non-perturbative corrections to an otherwise perturbative 
expansion; ii) a dispersion formula saturated by certain low-lying hadron 
resonances. Equating the two amounts to evaluating hadronic parameters in 
terms of the quark-level condensates. Despite certain conceptual problems
of "microscopic causality" encountered in the "matching" of two sides of the
equation, this method (QCD-SR) has proved very popular among a wide class of 
high energy phenomenologists, and has been continually refined over the 
years. A leisurely review of the state of the QCD-SR art on the quark 
structure of hardons, as well as its working on the problem of hadrons in 
nuclear matter (at finite temperature) is given by 
{\bf Leonard Kisslinger} [29] in this Book.

\subsection{Non-Perturbative Methods With QCD Features}

	The state of the art in this field is so diffuse that a more 
organized exposition is needed for such methods. To that end the attempts
at addressing the strong interaction problem in QCD may be divided into two
broad categories: i) soluble models designed to shed light on its general
features through exact calculations; and ii) effective Lagrangian methods 
for 4-fermion interactions, somewhat reminiscent of the Bethe "Second 
Principle" Theory of effective nucleon-nucleon interactions of the Fifties. 
Srivastava [17], as well as Kulshreshtha [18], in Part C of this Book, 
have already provided a flavour of the results to be expected from 
type (i) theories, using the method of LF-QFT. 
\par
	Type (ii) which deals with more realistic situations, albeit at the 
cost of some phenomenology, has a much wider literature to choose from. To 
do a semblance of justice to this field, this Book includes $two$ articles 
of this type, reviewing the methodology and working of such QFT-based 
approaches. The first one, by {\bf Vladimir Karmanov} [30], gives an 
in-depth review of covariant light-front (LF) dynamics, with applications 
to field theory and relativistic wave functions. The formalism is effectively
3D in content, which can be obtained by projecting the (4D) Bethe-Salpeter
amplitudes on the light-front plane, and although a reversal of steps is
not possible to reconstruct the 4D BS amplitude, the LF formalism still
represents a powerful alternative for solving  QFT problems. Karmanov [30]
also discusses some typical applications.          
     
\subsubsection{Markov-Yukawa Transversality Principle}

	The second article by {\bf Asoke Mitra} [31] offers a comparative 
view of the state of the art in several QFT approaches based on effective 
4-fermion interactions (including QCD features), of both 3D and 4D types
(Tamm-Dancoff, Bethe-Salpeter, Salpeter, quasi potentials, light-front).
In this context, attention is focussed on an important but somewhat less 
known principle called "Markov-Yukawa Transversality" ($MYTP$) which 
decrees that the interaction between the two (quark) constituents be 
$transverse$ to the composite (hadron) 4-momentum, by virtue of which the 
BSE kernel has an effective (albeit covariant) 3D support. As a result of 
this "Covariant Instantaneity" the starting 4D BSE is exactly reducible to 
a 3D form, and conversely the steps can be reversed so as to allow an $exact
reconstruction$ of the original 4D BSE in terms of 3D ingredients ! Thus
$MYTP$ allows an exact interlinkage between the 3D and 4D BSE forms, so 
that both forms can be used interchangeably, unlike most other approaches 
in the literature which employ either a 4D or a 3D form of the BS dynamics, 
but not both simultaneously.  
\par
	It might be of some historical interest to note that the Salpeter 
equation has a 3D structure stemming from its (instantaneous) kernel with
a 3D support, and therefore its original 4D form can be recovered a la 
$MYTP$ by reversing the steps, but this possibility had never been explored. 
This gap is now filled by $MYTP$ which provides a formally covariant basis 
for the instantaneous approximation. The same principle ($MYTP$) can also 
be generalized from covariant instantaneity to the covariant light-front. 
\par
	A fall-out of the 3D-4D interlinkage provided by $MYTP$ is that it
gives a $two-tier$ description: the 3D form for the hadron spectra which are 
$O(3)$-like; and the 4D form to address the transition amplitudes as 4D
loop integrals using standard (4D) Feynman rules. This Principle can be
easily incorporated in the usual framework of coupled Bethe-Salpeter and 
Schwinger-Dyson equations (BSE-SDE) stemming from a (chirally invariant) 
4-fermion Lagrangian with current quarks interacting via the full gluon 
propagator, so that the quark mass is acquired via the NJL-mechanism. And
the generalization from covariant instantaneity to the covariant light-front
helps remove certain problems of Lorentz mismatch of vertex functions that
arise in a 4D loop integral under the covariant instantaneity ansatz.  
These and other details are reviewed in the article by Mitra [31] which
also stresses a parallelism of treatment of $q{\bar q}$ and $qqq$ systems.
      
\subsection{The Harmonic Oscillator: A Powerful Bridge In QFT} 

No amount of literature on the impact of QFT in Physics would be complete 
without an exposure of the role of the Harmonic Oscillator (HO) in 
shaping Quantum Theory, as an integral part of this Book theme. It was 
therefore a matter of great satisfaction when {\bf Marcos Moshinsky} [32], 
who may be regarded as the "Father of the Harmonic Oscillator in Physics", 
agreed to contribute a perspective article on the HO theme. The only 
obstacle against a regular format for his Article was that he had only
recently written a comprehensive book on the subject [M. Moshinsky and
Yu.F.Smirnov, $The Harmonic Oscillator In Modern Physics$, (Harwood
Academic Press, the Netherlands, 1996)].  Nevertheless in his Article,
he has provided a comprehensive list of contents of his HO-book, which 
already offers a glimpse of the depth and range of physical problems 
(from the simplest quantum mechanical ones to the $n$-body Relativistic
Oscillator) that are amenable to the amazing powers of HO techniques
in tandem with the standard methods of Group Theory. In addition he has
reviewed some recent work of his on relativistic particles of arbitrary
spin in a $confining$ HO potential, with applications to Spectroscopy.    
       
\section{Conclusion: Foundations Of Quantum Theory}
   
We conclude this Introduction to the Book with an Article by 
{\bf Dipankar Home}[33] on the modern perspectives on the foundations of 
quantum mechanics (the predecessor of QFT), which are increasingly being 
scrutinized by relating them to precise experimental studies. In this 
Article, Home [33] picks two main issues: i) quantum measurement problem; 
ii) quantum non-locality, for a detailed exposition in a theory vs 
experiment scenario. He concludes with a quotation from John Bell: 
"It seems to me possible that the continuing anxiety about what quantum 
mechanics means or entails will lead to still more tricky experiments which 
will eventually find some soft spot." Translated to the QFT level, this 
looks like an appropriate conclusion for this Book as well.


\begin{thebibliography}{99} 
  
\bibitem{1}
 D.V.Shirkov: Evolution Of The Bogoliubov Renormalization Group 
\bibitem{2}
 S. Szpigel and R.J.Perry: The Similarity Renormalization Group
\bibitem{3}
 V.Novikov: Quantum Field Theory And The Standard Model - Bird's Eye View
\bibitem{4}
 P.K.Kabir: Broken Reflection Symmetries
\bibitem{5}
 D.Boyanovsky and H.J.de Vega: Dynamics Of Symmetry Breaking Out Of
 Equilibrium - From Condensed Matter To QCD And The Early Universe
\bibitem{6}
 V.N.Gribov (Orsay Lectures): (I) hep-ph/9403218; (II) hep-ph/9404332 ; 
 (III) hep-ph/9905285             
\bibitem{7}
 K.Nishijima and M.Chaichian: An Essay On Color Confinement  
\bibitem{8}
 R.Kaul: Topological Quantum Field Theories - A Meeting Ground For
 Physicists And Mathematicians
\bibitem{9}
 E.Witten: Quantum Field Theory And The Jones Polynomial
\bibitem{10}
 H.Banerjee: Chiral Anomalies In Field Theories
\bibitem{11}
 Wei-Min Zhang: Coherent States In Field Theory
\bibitem{12}
 N.Mukunda: Pancharatnam, Bargmann And Berry Phases - A Retrospective
\bibitem{13}
 J.Schechter and H.Weigel:  The Skyrme Model For Baryons  
bibitem{14}
 R.Ramanathan: Euclidean Methods In Quantum Field Theory
\bibitem{15}
 Ashoke Das:  Topics In Finite Temperature Field Theory 
\bibitem{16}
 B.M.Sodermark: Integrable Models And The Toda Lattice Hierarchy
\bibitem{17}
 P.P.Srivastava: Perspectives Of Light-Front Quantized Field Theory -
 Some New Results
\bibitem{18}
 D.S.Kulshreshtha: Gauge Symmetry In Chiral Electrodynamics
\bibitem{19}
 L.Lusanna:  Towards A Unified Description Of The Four Interactions
 In Terms Of Dirac-Bergmann Observables 
\bibitem{20}
 R.N.Mohapatra: Supersymmetry And Particle Physics
\bibitem{21}
 N.Sakai: Supersymmetry In Field Theory
\bibitem{22}
 W.Nahm: Conformal Field Theory: A Bridge Over Troubled Waters 
\bibitem{23}
 J.H.Schwarz: Superstring Theory - An Overview
\bibitem{24}
 J.Maharana: Recent Developments In String Theory
\bibitem{25}
 L.Bonora:  Yang-Mills Theory And Matrix String Theory  
\bibitem{26}
 Avinash Khare: Fractional Statistics And Chern-Simons Field Theory
 In $2+1$ Dimensions
\bibitem{27}
 R.Rajaraman: Chern Simons Field And Composite Bosons In The Quantum
 Hall System 
\bibitem{28}
 O.Pene: Hadrons From QCD - Achievements And Prospects
\bibitem{29}
 L.S.Kisslinger: QCD Sum Rules In Hadronic And Nuclear Physics
\bibitem{30}
 V.A.Karmanov: Light-Front Dynamics 
\bibitem{31}
 A.N.Mitra: 3D-4D Interlinkage Of Bethe-Salpeter Amplitudes - A Unified
 View Of $Q{\bar Q}$ And $QQQ$ Dynamics
\bibitem{32}
 M.Moshinsky:  The Harmonic Oscillator In Quantum Theory - A Powerful
 Bridge In Physics  
\bibitem{33}
 D.Home: Modern Perspectives On Foundations Of Quantum Mechanics       

\end{thebibliography}
\end{document}